%% file: main.tex
\documentclass{article}

\usepackage{arxiv}

\usepackage[utf8]{inputenc} %
\usepackage[T1]{fontenc}    %
\usepackage{hyperref}       %
\usepackage{url}            %
\usepackage{booktabs}       %
\usepackage{amsfonts}       %
\usepackage{nicefrac}       %
\usepackage{microtype}      %
\usepackage{lipsum}		%
\usepackage{graphicx}
\usepackage{natbib}
\usepackage{doi}

\input{my_package}

\title{Detecting Inactive Cyberwarriors from Online Forums\thanks{This work is partially supported by the National Science and Technology
Council of Taiwan under grant 110-2222-E-008-005-MY3.}}

\date{} 					%

\author{{Ruei-Yuan Wang, Hung-Hsuan Chen}\\
	Department of Computer Science and Information Engineering\\
	National Central University\\
	Taoyuan, Taiwan \\
	\texttt{zx57680@gmail.com, hhchen1105@acm.org} \\
}

\date{}

\hypersetup{
pdftitle={Detecting Inactive Cyberwarriors from Online Forums},
pdfsubject={cs.AI, cs.LG},
pdfauthor={Ruei-Yuan Wang, Hung-Hsuan Chen},
pdfkeywords={cyber attack, graphical neural network, forum, spammer, netizen, information warfare, media framing, filter bubble, cyberwarrior},
}

\begin{document}
\maketitle

\begin{abstract}

The proliferation of misinformation has emerged as a new form of warfare in the information age. This type of warfare involves cyberwarriors, who deliberately propagate messages aimed at defaming opponents or fostering unity among allies. In this study, we investigate the level of activity exhibited by cyberwarriors within a large online forum, and remarkably, we discover that only a minute fraction of cyberwarriors are active users. Surprisingly, despite their expected role of actively disseminating misinformation, cyberwarriors remain predominantly silent during peacetime and only spring into action when necessary. Moreover, we analyze the challenges associated with identifying cyberwarriors and provide evidence that detecting inactive cyberwarriors is considerably more challenging than identifying their active counterparts. Finally, we discuss potential methodologies to more effectively identify cyberwarriors during their inactive phases, offering insights into better capturing their presence and actions.  The experimental code is released for reproducibility: \url{https://github.com/Ryaninthegame/Detect-Inactive-Spammers-on-PTT}.

\end{abstract}

\keywords{cyber attack \and graphical neural network \and forum \and spammer \and netizen \and information warfare \and media framing \and filter bubble \and cyberwarrior}

\section{Introduction} \label{sec:intro}

Social media has emerged as a crucial platform for information sharing, leading politicians, political parties, and governments to enlist the services of public relations (PR) companies and social media curators to bolster their online reputations. Regrettably, these PR firms occasionally engage in the deliberate dissemination of plausible but potentially incorrect or partially accurate statements on the Internet, employing techniques such as spin control or media framing. A prominent example of this phenomenon is Russia's interference in the 2016 United States election, with propaganda estimated to have reached 126 million Facebook users and over 20 million Instagram users~\citep{diresta2019tactics}.

Online propaganda typically relies on a multitude of user accounts to spread information and create a false impression of the formation of public opinion. These accounts, referred to as "cyberwarriors" in this paper, can be generated automatically or purchased at an affordable cost online. For example, a Chinese government document in 2021 reveals that accessing hundreds of active Facebook and Twitter accounts costs 5000 RMB per month, approximately 710 USD~\citep{xiao2021buying}.

The detection of cyberwarriors plays a pivotal role in combating the propagation of fake information. Previous studies have explored the behaviors of suspicious accounts and spammers on various platforms, proposing methodologies to detect them~\citep{hu2014online, gao2010detecting, benevenuto2010detecting, tan2012spammer, lu2013simultaneously, hu13social}. However, most of these studies focus on highly active users who exhibit extensive engagement on the platform, such as leaving comments, sharing photos, and initiating discussions. Apparently, detecting active spammers based on abundant activity logs is comparatively straightforward.

\begin{table}[tb]
\centering
\caption{A comparison of the AUPRC scores of detecting active and inactive cyberwarriors using different machine learning models.  The results show that detecting inactive cyberwarriors is much more challenging.}
\begin{tabular}{c||ccc}
\hline
& active users & inactive users & diff \\ \hline\hline
XGBoost & $0.8892$ & $0.5157$ & $0.3735$ \\
LightGBM & $0.7421$ & $0.4888$ & $0.2533$ \\
Random Forest & $0.8317$ & $0.5147$ & $0.3163$ \\ \hline
\end{tabular}
\label{tab:active-inactive-cmp}
\end{table}

Perhaps to the surprise of many people, although the mission of a cyberwarrior is to disseminate messages, a cyberwarrior account may remain inactive for an extended period before disseminating misleading posts~\citep{lin193things}. Consequently, active cyberwarriors make up only a tiny proportion of the total cyberwarriors. As a result, identifying inactive cyberwarriors may pose a significantly greater challenge. To validate this point, we conducted a preliminary study demonstrating the ease of detecting spammers among active users using supervised learning techniques and the difficulty of detecting inactive spammers. As illustrated in Table~\ref{tab:active-inactive-cmp}, when applying some of the most successful machine learning models (XGBoost, LightGBM, and Random Forest) to detect spammers among active users, we achieve decent scores. However, these numbers decrease significantly when targeting inactive users, with an average drop of over $30\%$ in the area under the precision-recall curve (AUPRC).

In this paper, our aim is to quantify a user's level of activeness and focus on identifying abnormal accounts among inactive users. Since inactive users provide limited activity logs as features, we enhance the available clues by incorporating social information from two perspectives. First, we use data from inactive users' connected accounts to generate social-related features. Second, we employ graph neural networks (GNNs) as training models to capture the relationships between different accounts. Consequently, even if an account exhibits limited activity logs, we can leverage information from its neighboring accounts to detect its status (normal or spammer). Our findings indicate that these simple strategies significantly improve the effectiveness of discovering inactive spammers while also slightly enhancing the detection of active spammers.

In summary, this paper makes the following contributions.

\begin{itemize}
\item We demonstrate that detecting spammers among inactive users is considerably more challenging than among active users. Additionally, we highlight that a substantial number of spammers are inactive users, which has not received significant attention in previous studies that primarily focus on active users.
\item We introduce social-related features and employ graph neural network models to leverage information from an account's neighboring accounts. Through comprehensive experiments, we demonstrate that these simple strategies improve the detection of inactive and active spammers compared to other baseline methods.
\item For reproducibility purposes, we release the experimental dataset and the accompanying code. In addition, the dataset can serve as a valuable benchmark for spammer detection, as administrators of a large forum have manually labeled the spammers in our dataset.
\end{itemize}

The rest of the paper is organized as follows. Section~\ref{sec:rel-work} reviews previous studies on spammer detection. In Section~\ref{sec:ptt}, we present the statistics and features of our studied forum. In Section~\ref{sec:exp}, we analyze the behaviors of active and inactive spammers and compare the challenges associated with their detection. Finally, we conclude and discuss our work in Section~\ref{sec:disc}.

\section{Related Work} \label{sec:rel-work}

Detecting abnormal accounts has been a topic of extensive research, with various approaches and techniques employed to establish the relationship between account features and their classification as normal or abnormal. Machine learning models have proven to be valuable tools in this regard~\citep{hu2014online, gao2010detecting, benevenuto2010detecting, tan2012spammer, lu2013simultaneously, hu13social}. These models leverage relevant information, such as an account's public profile and behaviors, to identify patterns indicative of abnormal activity. Although these machine learning methods have demonstrated promising performance in determining the status of active accounts, they often encounter difficulties when dealing with less active accounts, as evidenced by our preliminary study in Table~\ref{tab:active-inactive-cmp}. The limited information left by inactive accounts poses challenges for feature extraction and classification, leading to suboptimal performance in detecting such accounts.

Graphs are a natural choice for modeling relationships due to their ability to represent complex interactions and connections between entities in a visually intuitive and versatile manner~\citep{chen2011collabseer, chen2013csseer, chen2015ascos++, chen2012discovering}. In recent years, graph neural networks have received significant attention due to their remarkable performance in various domains, including biomedical research, social network analysis, and abnormal sensor detection~\citep{zhou2020graph, fan2019graph, wu23detecting}. Exploiting the inherent graph structure of social networks to detect abnormal accounts becomes a natural choice. Consequently, researchers have leveraged graph-based approaches~\citep{wang2011review} and, more recently, graph neural networks (GNNs)~\citep{yang2022rosgas, shi2022h2, huang2022auc} to model social networks and make predictions. These techniques can effectively identify suspicious patterns and uncover abnormal behavior by capturing relational information among accounts. However, despite the progress made in this field, detecting abnormal accounts remains challenging, particularly when dealing with less active or inactive accounts. The complexities that arise from the limited availability of information and the evolving nature of suspicious behavior necessitate further research and the development of advanced techniques to enhance detection accuracy and robustness.

\section{The PTT Forum} \label{sec:ptt}

We collect the experimental dataset from the PTT forum. In this section, we provide a comprehensive introduction to the PTT forum, including its background, statistics, and noteworthy features, to familiarize readers with the platform.

\subsection{Introduction of PTT}

PTT, established in 1995, has emerged as one of the largest and most influential forums in Taiwan. With a massive user base that exceeds 1.5 million registered accounts and encompasses over 20,000 discussion boards covering a wide range of topics, PTT serves as a vibrant platform where users actively engage in discussions, sharing insights, opinions, and experiences~\citep{wikipediaptt22}. The popularity of the forum is evidenced by the staggering volume of user-generated content, with an average daily production of more than 20,000 articles and over 500,000 comments.

The PTT forum caters to the diverse interests of Taiwanese citizens, providing an avenue for discussions on a myriad of subjects, including shopping experiences, celebrity gossip, news updates, religions, movies, life goals, and notably critical societal events. In particular, the platform has played an important role in facilitating discussions during key historical moments in Taiwan in the last decades. For instance, during the Sunflower Student Movement in 2014, which involved a three-week occupation of Taiwan's Legislative Yuan\footnote{Logislative Yuan is the unicameral legislature of Taiwan, similar to UK Parliament and US Congress.} by civic groups and students, a single discussion board on PTT witnessed the simultaneous presence of over 100,000 users, which further encouraged more citizens to join the movement.  This event demonstrates the forum's ability to mobilize individuals and foster engagement. Similarly, during the 2016 presidential election and the 2018 city mayor election, PTT attracted similar numbers of users concurrently visiting a discussion board, further highlighting its relevance and impact in shaping public discourse.

Given the substantial influence of PTT, various entities, such as journalists, politicians, political parties, and the entertainment industry, actively monitor the platform to gauge public opinion. In particular, politicians recognize the importance of securing votes, particularly from the young and middle-aged demographics, by leveraging PTT as a battleground to connect with potential supporters and address their concerns.

PTT stands out among other online forums due to its distinctive features and mechanisms. One notable feature is its commenting system, where users can express their sentiment towards an article through options such as liking, disliking, or remaining neutral, accompanied by a 45-character comment. Moreover, articles that receive a significant number of likes or dislikes are visually highlighted with special colored symbols, capturing users' attention and potentially triggering further engagement. This feedback loop reinforces the amplification of likes or dislikes and subsequently increases the visibility of such articles, leading to increased exposure and potential impact.

However, with the substantial influence and visibility of PTT, there have been instances where politicians, political parties, and public relations (PR) firms resort to disseminating disinformation on the platform for various purposes, including media framing, attacking opponents, or self-promotion. The unique highlighting system introduced earlier serves as a motivation for individuals with specific agendas to mobilize accounts and accumulate a large number of likes or dislikes on selected articles in a short period of time, aiming to generate further attention around these topics~\citep{nguyen2021learning}. These dynamics present challenges in distinguishing between genuine user participation and orchestrated manipulations, underscoring the need for robust detection mechanisms.

\subsection{Experimental Data Collection on PTT}

To conduct our research, we collected experimental data from the PTT forum, focusing on a specific time period and a subset of accounts associated with suspicious activities. From March 2019 to December 2019, PTT officially announced $7,581$ accounts as spammers, primarily suspected of attempting to influence the city mayor elections of six major cities in Taiwan in November 2018 and the upcoming presidential election in March 2020. Out of these $7,581$ accounts, $4,918$ of them have at least one activity record related to article posting or commenting. However, it is worth noting that most of these $4,918$ accounts exhibit minimal activity, with up to $92\%$ of them having no more than 0.18 activities per day, indicating a high degree of inactivity. Consequently, relying solely on the activity logs of these accounts to detect whether they are spammers or regular users, as suggested by previous studies, may not yield optimal results.

To capture the relevant data for our analysis, we crawled the articles from July 1, 2018, to December 29, 2019, based on the following considerations. First, the PTT announced the first batch of suspicious accounts in March 2019, approximately four months after the city mayors' election on November 24, 2018. Given this timeline, we assume that these accounts began their actions approximately six months prior to the election. Therefore, we started our data collection on July 1, 2018, to include the crucial period leading up to the election. Secondly, the PTT announced its last batch of suspicious accounts and suspended their posting and commenting permissions on December 29, 2019. Consequently, we set this date as the final crawling day to ensure comprehensive coverage of relevant data.

After crawling the articles and comments, we discovered that the total number of associated accounts (that is, including the authors and commentors) exceeded $200,000$, which would require substantial memory space, particularly when employing graph neural networks (details of which will be introduced in Section~\ref{sec:exp}). To manage the dataset more effectively, we further pruned the articles based on specific criteria. First, we included only articles with at least 90 comments, ensuring a reasonable level of engagement for comprehensive analysis. Second, if the associated accounts of an article contained fewer than three spammers, we excluded the article, focusing on those articles where suspicious activities were more prevalent. Finally, we include a maximum of 80 commentors for the remaining articles. Specifically, if the number of spammers associated with an article was less than 80, we included all the spammers; we included regular users in chronological order until we reached a total of 80 accounts. On the contrary, if more than 80 spammers were associated with an article, we selected the earliest 80 spammers while excluding regular user accounts. Following these criteria, we collected a dataset consisting of $44,602$ user accounts, with $912$ of them identified as spammers by PTT administrators.

All subsequent experiments and analyses presented in this study are based on the pruned dataset obtained after the selection process.

\section{Analysis} \label{sec:exp}

This section presents the empirical activeness scores of cyberwarriors and compares the effectiveness of different models to detect them. It provides insights into the activity levels of spammers and explores the performance of various algorithms in identifying them.

\subsection{Most spammers are less active than normal users} \label{sec:spammers-are-inactive}

\begin{table*}[tb]
\centering
\caption{The number of normal users and spammers for each group.  The symbol $[p, q)$ refers to the percentile of active value $r$ in the range: $p \le r < q$.}
\label{tab:active-value-user-percentile}
\resizebox{\textwidth}{!}{%
\begin{tabular}{ccccccccc}
    \toprule
    Group & \makecell{Percentile of\\ active value}   & \makecell{Active value}     & \makecell{\# normal\\ accounts} & \makecell{CDF of normal accounts \\ (a)} & \makecell{\# spammers} & \makecell{CDF of spammers \\ (b)} & (b) $-$ (a) \\
    \midrule
    $G_1$ & [0\%, 10\%) &  0-18      &   4112  & $9\%$ &    222  & $24\%$ & $15\%$ \\
    $G_2$ & [10\%, 20\%)  &  19-45     &   4418  & $20\%$  &  163  & $42\%$ & $22\%$ \\
    $G_3$ & [20\%, 30\%)  &  46-84     &   4508  & $30\%$ &    86  & $52\%$ & $22\%$ \\
    $G_4$ & [30\%, 40\%)  &  85-135    &   4223  & $40\%$ &    59  & $58\%$ & $18\%$ \\
    $G_5$ & [40\%, 50\%)  &  136-211   &   4453  & $50\%$ & 57  & $64\%$ & $14\%$ \\
    $G_6$ & [50\%, 60\%)  &  212-315   &   4096  & $59\%$ & 76 & $73\%$ & $14\%$ \\ 
    $G_7$ & [60\%, 70\%)  &  316-494   &   4320  & $69\%$ & 112  & $85\%$ & $16\%$ \\
    $G_8$ & [70\%, 80\%)  &  495-817   &   4368  & $79\%$ & 67  & $92\%$ & $13\%$ \\
    $G_9$ & [80\%, 90\%)  &  818-1663  &   4638  & $90\%$ & 51  & $98\%$ & $8\%$ \\
    $G_{10}$ & [90\%, 100\%] &  $\ge 1664$     & 4554  & $100\%$ & 19  & $100\%$ & $0\%$ \\
    \bottomrule
\end{tabular}
}
\end{table*}

We define the degree of activeness of an account by considering the average number of daily articles and comments. To assess the activity levels of the collected users, we calculate the active value for each user and categorize them into 10 groups, denoted $G_1$ to $G_{10}$. Each group $G_i$ contains users whose active values fall within the $(i-1)$th percentile and the $i$th percentile among all users.

Table~\ref{tab:active-value-user-percentile} presents the number of normal and spammer accounts in each group $G_i$. As evident from the column "\# normal accounts" and the column "CDF of normal accounts", the number of normal accounts remains relatively consistent across the groups. However, the activeness values of spammers exhibit a significant skew. As indicated in the last column of Table~\ref{tab:active-value-user-percentile}, the cumulative distribution function (CDF) of spammers for each row consistently exceeds the CDF of regular accounts. This implies that, compared to normal users, most spammers exhibit lower activity levels.

Since cyberwarriors are expected to disseminate information, it may be argued that cyberwarriors should demonstrate higher levels of activity. Thus, our empirical observation -- spammers are typically less active during non-conflict periods -- may be a surprise to many people.  However, we found that previous research on Twitter accounts aligns with our findings and supports the claim that cyberwarriors often exhibit extended periods of inactivity during peacetime and only engage in extensive posting when necessary~\citep{lin193things}.  

\subsection{Supervised learning is successful in detecting active spammers, but not inactive spammers}

\begin{table}[tb]
\centering
\caption{The AUPRC scores of various non-GNN models (without social features).  We repeat each experiment 10 times and report the mean $\pm$ standard deviation.}
\label{tab:nongnn-auprc-cmp}
\begin{tabular}{c||ccc}
    \toprule
     & [0\%, 10\%)      & [10\%, 20\%)     & [80\%, 100\%]     \\
    \midrule
    XGBoost              & $0.52 \pm 0.01$ & $0.48 \pm 0.03$ & $0.89 \pm 0.01$     \\
    LightGBM             & $0.49 \pm 0.02$ & $0.40 \pm 0.04$ & $0.74 \pm 0.02$     \\
    Random Forest         & $0.51 \pm 0.03$ & $0.27 \pm 0.02$ & $0.83 \pm 0.02$  \\
    Fully Connected      & $0.35 \pm 0.06$ & $0.38 \pm 0.05$ & $0.75 \pm 0.03$  \\ 
    ConvNet     & $0.17 \pm 0.06$ & $0.26 \pm 0.14$ & $0.80 \pm 0.33$  \\
    Soft Voting~\citep{nguyen2021learning} & $0.40 \pm 0.01$ & $0.43 \pm 0.01$ & $0.76 \pm 0.01$  \\
    Hard Voting~\citep{nguyen2021learning} & $0.43 \pm 0.02$ & $0.47 \pm 0.02$ & $0.70 \pm 0.03$   \\
    Stacking~\citep{nguyen2021learning}  & $0.42 \pm 0.01$ & $0.47 \pm 0.03$ & $0.67 \pm 0.01$   \\
    \bottomrule
\end{tabular}
\end{table}

Given that spammers are generally less active than normal users, detecting them may pose a greater challenge for algorithms because of the limited clues they leave behind.

To validate this conjecture, we selected various algorithms and tested their effectiveness in identifying active and inactive spammers.  The algorithms include two popular algorithms known for their success in Kaggle competitions (XGBoost and LightGBM), deep learning models such as fully connected networks and convolutional neural networks (ConvNet), and recently proposed approaches for spammer detection for PTT, namely soft voting, hard voting, and stacking ensemble~\citep{nguyen2021learning}.  For each account $a$, we considered three features. First, we computed the average popularity of a user's associated articles (i.e., the account $a$'s  posted or commented articles).  In particular, we computed the total number of comments for all $m$ articles and divided by $m$.  Second, we calculated the average sentiment of comments about articles by subtracting the number of dislikes from the number of likes for each of the $m$ articles and computing the average. These two features were included because previous studies indicate that spammers often generate many comments on selected articles to increase their visibility. Lastly, we incorporated the active period of an account as the third feature.

To evaluate the performance, we used the area under the precision-recall curve (AUPRC) as the metric. Given the highly imbalanced nature of our data set, with a percentage of spammers ranging from $0.4\%$ to $5\%$ in each group (as shown in Table~\ref{tab:active-value-user-percentile}), AUPRC was considered more appropriate than the area under the receiver operating characteristic curve (AUROC). AUROC tends to overstate the performance of a classifier when the positive class is the minority class, potentially leading to misleading results~\citep{boyd12unachievable, cook2020consult}. In contrast, AUPRC is suitable for scenarios where the positive class is of interest and represents the minority, as it accounts for precision and recall without considering true negatives (i.e., the negative instances that are predicted as negative by a model).

Table~\ref{tab:nongnn-auprc-cmp} reports the AUPRC scores of the selected algorithms for three groups based on users' activeness values: $[0, 10\%)$, $[10\%, 20\%)$, and $[80\%, 100\%]$. The results reveal that as the activeness value increases (i.e., in the $[80\%, 100\%]$ group), the improved scores of the average AUPRC range from $20\%$ to $63\%$ compared to the users in the $[0\%, 10\%)$ or $[10\%, 20\%)$ groups. This finding aligns with our hypothesis that supervised learning algorithms excel at detecting active cyberwarriors. However, identifying inactive cyberwarriors is significantly more challenging. Unfortunately, most cyberwarriors exhibit low activity during peacetime, making it possible to identify them only when they engage in aggressive posting and sharing of articles.

\subsection{Social information helps discover inactive spammers}

This section demonstrates that integrating social information can enhance the identification of inactive spammers. We explore two perspectives for incorporating social information: utilizing graphical neural network (GNN) models and designing specialized social features. Our experimental results validate the effectiveness of both approaches.

\subsubsection{GNN models}

\figwidth{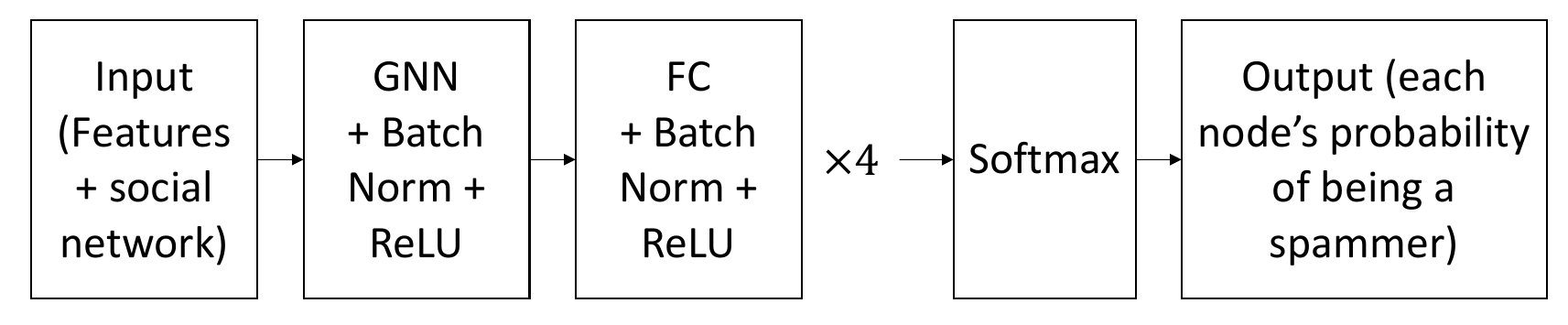}{The structure of the GNN-based models.  The GNN is either GCN, TAGCN, or GAT.}{fig:gnn-structure}{.7\textwidth}

\begin{table}[tb]
\centering
\caption{The AUPRC scores of various GNN models and the best non-GNN model (without social features).  We repeat each experiment 10 times and report the mean $\pm$ standard deviation.}
\label{tab:gnn-auprc-cmp}
\begin{tabular}{c||ccc}
    \toprule
     & [0\%, 10\%)      & [10\%, 20\%)     & [80\%, 100\%]     \\
    \midrule
XGBoost              & $0.52 \pm 0.01$ & $0.48 \pm 0.03$ & $0.89 \pm 0.01~\dag$     \\
    \midrule
    GCN                  & $\boldsymbol{0.66} \pm 0.18$ & $0.38 \pm 0.13$ & $0.72 \pm 0.07$   \\
    TAGCN ($K=1$)           & $\boldsymbol{0.64} \pm 0.04$ & $\boldsymbol{0.79} \pm 0.06$ & $\boldsymbol{0.89} \pm 0.07~\dag$   \\
    TAGCN ($K=2$)           & $\boldsymbol{0.68} \pm 0.02$ & $\boldsymbol{0.84} \pm 0.05~\dag$ & $\boldsymbol{0.89} \pm 0.08~\dag$   \\
    TAGCN ($K=3$)           & $\boldsymbol{0.71} \pm 0.04~\dag$ & $\boldsymbol{0.80} \pm 0.07$ & $\boldsymbol{0.89} \pm 0.06~\dag$   \\
    GAT                  & $\boldsymbol{0.62} \pm 0.09$ & $\boldsymbol{0.77} \pm 0.05$ & $\boldsymbol{0.89} \pm 0.06~\dag$    \\  
    \bottomrule
\end{tabular}
\end{table}

This section introduces GNN-based models and explains how we construct a social network. Leveraging the social information embedded in the network, models can potentially extract valuable insights from neighboring accounts, even when an inactive account provides limited activity logs.

We consider each account as a node in a graph, connecting two nodes with an edge if the corresponding accounts co-appear in an article (either as commentors or with one as the poster and the other as a commentor). To represent the graph, we generate an adjacency matrix $A=[a_{i,j}]_{i,j=1,\dots,n}$, where $n$ denotes the number of nodes, and $a_{i,j} = 1$ if there exists an edge connecting nodes $i$ and $j$, and 0 otherwise.

We selected three representative GNN models as part of our learning framework: graph convolutional networks (GCN)~\citep{welling2016semi}, topology adaptive graph convolutional networks (TAGCN)~\citep{du2017topology}, and graph attention network (GAT)~\citep{velickovic18graph}. These GNNs incorporate information from neighboring nodes into each node $i$ through recursive information propagation, thereby fusing the neighboring information with node $i$'s information. The differences among these models lie in the range of the neighboring area and the mechanism used to integrate information. Figure~\ref{fig:gnn-structure} provides an overview of the structure of the neural network with GNN models.

Table~\ref{tab:gnn-auprc-cmp} compares the best non-GNN model (XGBoost) with GNN-based models. Most models perform satisfactorily in detecting spammers from active accounts, as indicated in the last column. However, when targeting less active accounts, most GNN-based models outperform the best non-GNN model. We highlight the best performing model for each column using the $\dag$ symbol.  If a GNN model performs better than or at least as well as XGBoost, we highlight it in bold.

\subsubsection{Social-related features}

\begin{table}[tb]
\caption{The average suspect value for the users of different degrees of activeness}
\label{tab:suspect-stats}
\centering
\begin{tabular}{ccc}
\toprule
\makecell{Percentile of\\ active value} & \# spammers & \makecell{Average\\ suspect value} \\ \midrule
$[0\%-10\%)$ & 222 & 16.487 \\
$[10\%-20\%)$ & 163 & 4.682 \\
$[20\%-30\%)$ & 86 & 3.577 \\
$[30\%-40\%)$ & 59 & 2.778 \\
$[40\%-50\%)$ & 57 & 1.931 \\
$[50\%-60\%)$ & 76 & 1.579 \\
$[60\%-70\%)$ & 112 & 1.126 \\
$[70\%-80\%)$ & 67 & 0.784 \\
$[80\%-90\%)$ & 51 & 0.511 \\
$[90\%-100\%]$ & 19 & 0.123 \\ \midrule
$[0\% - 100\%]$ & 912 & 5.899 \\    
\bottomrule
\end{tabular}
\end{table}

\begin{table*}[tb]
\centering
\caption{A comparison of various models (including social features) in terms of the AUPRC score.  We repeat each experiment 10 times and report the mean $\pm$ standard deviation. We highlight the winner of non-GNN-based models in bold.  We highlight the GNN-based models if they outperform the best non-GNN-based models.}
\label{tab:auprc-cmp-include-suspicious-value}
\resizebox{\textwidth}{!}{%
\begin{tabular}{c|c|ccc|c}
    \toprule
 Type & Model & [0\%, 10\%)      & [10\%, 20\%)     & [80\%, 100\%]  & [0\%, 100\%]   \\
    \midrule
 \multirow{8}{*}{\makecell{Non-GNN-based models\\(including social features)}}&  XGBoost              & $0.83 \pm 0.01$ & $\boldsymbol{0.74} \pm 0.03$ & $\boldsymbol{0.90} \pm 0.02$ & $\boldsymbol{0.86} \pm 0.00$\\
 &  LightGBM             & $\boldsymbol{0.86} \pm 0.02$ & $0.72 \pm 0.05$ & $0.88 \pm 0.02$ & $0.82 \pm 0.00$ \\
 &  Random Forest         & $0.85 \pm 0.01$ & $0.56 \pm 0.05$ & $0.85 \pm 0.02$ & $0.79 \pm 0.00$ \\
 &  Fully Connected      & $0.53 \pm 0.07$ & $0.51 \pm 0.06$ & $0.76 \pm 0.05$ & $0.64 \pm 0.04$ \\ 
 &  ConvNet         & $0.43 \pm 0.09$ & $0.68 \pm 0.07$ & $0.83 \pm 0.04$ & $0.66 \pm 0.06$ \\
 &  Soft Voting~\citep{nguyen2021learning} & $0.69 \pm 0.00$ & $0.56 \pm 0.01$ & $0.76 \pm 0.01$ & $0.72 \pm 0.00$  \\
 &  Hard Voting~\citep{nguyen2021learning} & $0.67 \pm 0.01$ & $0.63 \pm 0.02$ & $0.70 \pm 0.03$ & $0.74 \pm 0.01$ \\
 &  Stacking~\citep{nguyen2021learning} & $0.54 \pm 0.02$ & $0.56 \pm 0.03$ & $0.67 \pm 0.01$ & $0.69 \pm 0.02$ \\
  \midrule
 \multirow{5}{*}{\makecell{GNN-based models\\(including social features)}} & GCN                  & $0.62 \pm 0.08$ & $0.52 \pm 0.05$ & $0.83 \pm 0.08$ & $0.69 \pm 0.03$  \\
 & TAGCN ($K=1$)           & $0.79 \pm 0.03$ & $\boldsymbol{0.97} \pm 0.05$ & $\boldsymbol{0.99} \pm 0.04$ & $\boldsymbol{0.92} \pm 0.01$  \\
 & TAGCN ($K=2$)           & $0.82 \pm 0.03$ & $\boldsymbol{0.98} \pm 0.02$ & $\boldsymbol{0.99} \pm 0.03$ & $\boldsymbol{0.93} \pm 0.02$ \\
 & TAGCN ($K=3$)           & $0.85 \pm 0.02$ & $\boldsymbol{0.98} \pm 0.03$ & $\boldsymbol{0.98} \pm 0.01$ & $\boldsymbol{0.94} \pm 0.01$ \\
 & GAT                  & $0.73 \pm 0.06$ & $\boldsymbol{0.91} \pm 0.06$ & $\boldsymbol{0.92} \pm 0.07$ & $\boldsymbol{0.87} \pm 0.05$ \\  
    \bottomrule
\end{tabular}
}
\end{table*}

The previous section illustrates that integrating social information helps identify less active spammers. This discovery led us to hypothesize that by designing social-related features, we could potentially assist non-GNN models in detecting less active cyberwarriors.

We introduce a new feature, the suspect value $s_i$, for each account $i$. The suspect value $s_i$ is defined as the ratio of the number of times user $i$ co-occurs with any spammer in an article to user $i$'s activeness value, as expressed by Equation~\ref{eq:suspect-value}.

\begin{equation} \label{eq:suspect-value}
    s_i = \frac{\sum_{\forall p \in \mathcal{A}_i}I(p \in \mathcal{P}^{(\textrm{spammer})})}{a_i},
\end{equation}
where $a_i$ represents the activeness value of user $i$, $\mathcal{A}_i$ denotes the set of articles associated with user $i$ (i.e., the set of articles in which user $i$ has either posted or commented), $\mathcal{P}^{(\textrm{spammer})}$ returns the set of articles posted or commented on by spammers, and $I$ denotes the indicator function, such that $I(x) = 1$ if $x$ is true and $0$ otherwise. 

Table~\ref{tab:suspect-stats} presents the average suspect values for different ranges of activeness values. The results reveal a clear relationship between a user's activeness and their suspect value: users with lower activity levels tend to connect with more spammer accounts. This finding supports our earlier observation in Section~\ref{sec:spammers-are-inactive} that spammers exhibit less activity. Specifically, we find that inactive accounts tend to have more connections to spammers, which could indicate suspicious behavior.

Table~\ref{tab:auprc-cmp-include-suspicious-value} reports the AUPRC scores of both non-GNN models and GNN-based models to detect cyberwarriors in different degrees of activeness, incorporating the suspect value as a feature. This social feature enhances the identification of cyberwarriors for both non-GNN and GNN-based models (referring to Table~\ref{tab:nongnn-auprc-cmp} and Table~\ref{tab:gnn-auprc-cmp}). Additionally, the suspect value feature proves particularly helpful in identifying spammers from the inactive user groups, as exemplified by LightGBM's AUPRC increasing from $0.49$ to a remarkable $0.86$ for the most inactive group of users.

\subsection{F1 scores When Claiming the Top-$k$ Suspicious Users as cyberwarriors}

\begin{table*}[tb]
\centering
\caption{A comparison of various models (including social features) in terms of the F1 score.  We highlight the winner of non-GNN-based models in bold.  We highlight a GNN-based model if its result outperforms the best non-GNN model.}
\label{tab:f1-score-cmp}
\begin{tabular}{c|c|cccc}
\toprule
Type & Model & $k=100$            & $k=200$            & $k=300$            & $k=400$            \\ \midrule
\multirow{8}{*}{\makecell{Non-GNN-based models\\(including social features)}} & LightGBM      & 0.6078          & \textbf{0.7729} & 0.6832          & \textbf{0.5969} \\
& XGBoost       & \textbf{0.6431} & 0.7676          & \textbf{0.6915} & 0.5866          \\
& Random Forest & 0.6042          & 0.7598           & 0.6811          & 0.5849          \\
& ConvNet       & 0.6289          & 0.7154          & 0.6336          & 0.5523          \\
& FC            & 0.4382          & 0.6162          & 0.5880          & 0.5352          \\
& Ensemble     & 0.1594          & 0.2325          & 0.2193          & 0.2096          \\
& Soft Voting   & 0.1838          & 0.3148          & 0.4208          & 0.5092          \\
& Hard Voting   & 0.1640          & 0.2842          & 0.3977          & 0.488           \\
\midrule
\multirow{5}{*}{\makecell{GNN-based models\\(including social features)}} & GATC        & 0.4382          & 0.6319          & \textbf{0.6916} & \textbf{0.6038} \\
& GCN          & \textbf{0.6573} & 0.6632          & 0.5549          & 0.4700          \\
& TAGCN ($K=1$)   & \textbf{0.6926} & \textbf{0.8564} & 0.6873          & 0.5695          \\
& TAGCN ($K=2$)    & \textbf{0.6997} & \textbf{0.8669} & \textbf{0.7122} & \textbf{0.5970} \\
& TAGCN ($K=3$)    & \textbf{0.7067} & \textbf{0.8721} & \textbf{0.7164} & \textbf{0.6072}  \\ \bottomrule
\end{tabular}
\end{table*}

After a model predicts the probability of an account being abnormal for each user, practical verification from administrators is still necessary. Therefore, a two-step procedure can be employed to determine suspicious accounts in practice. The procedure involves ranking all users based on their predicted suspiciousness using a prediction model, followed by manual examination of the top-$k$ most suspicious accounts by administrators or guardians. The value of $k$ is determined based on the available manpower, allowing for the verification of suspicious accounts with minimal labor costs.

To evaluate the effectiveness of the aforementioned two-step approach using different prediction models, we compute the $F1$-at-$k$ ($F1@k$) scores for varying values of $k$. The $F1@k$ score is defined in Equation~\ref{eq:f1-at-k}.

\begin{equation} \label{eq:f1-at-k}
F1@k = 2 \times \frac{p@k \times r@k}{p@k + r@k},
\end{equation}
where $p@k$ and $r@k$ are precision-at-$k$ and recall-at-$k$, defined by Equation~\ref{eq:p-at-k} and Equation~\ref{eq:r-at-k}, respectively.

\begin{equation} \label{eq:p-at-k}
    p@k = \frac{\textrm{number of abnormal accounts in top }k}{k}
\end{equation}

\begin{equation} \label{eq:r-at-k}
    r@k = \frac{\textrm{number of abnormal accounts in top }k}{\textrm{total number of abnormal accounts}}
\end{equation}

The $F1@k$ score extends the standard $F1$ measure to evaluate a ranked list by considering the top-$k$ predictions. It provides a comprehensive assessment by integrating both $p@k$ and $r@k$. The precision-at-$k$ ($p@k$) measures the proportion of abnormal accounts among the top-$k$ suspicious accounts, indicating the accuracy of the model's predictions within the top-$k$ positions. On the other hand, recall-at-$k$ ($r@k$) focuses on the completeness of predictions by evaluating how many abnormal accounts are included among the top-$k$ positions. It indicates the model's ability to identify and retrieve relevant items from the entire set. The harmonic mean of $p@k$ and $r@k$ is used to compute the $F1@k$ score, ensuring that both precision and recall contribute to the final evaluation.

Table~\ref{tab:f1-score-cmp} presents the $F1@k$ scores for various models at different values of $k$. The results demonstrate that LightGBM and XGBoost remain the top-performing models of non-GNN-based approaches. However, GNN models consistently outperform the best non-GNN models on $F1@k$ for different $k$ values. Therefore, when employing the two-step human-machine cooperation strategy described above, utilizing GNN-based models with social features remains a favorable option.

\section{Discussion} \label{sec:disc}

This paper contributes to understanding spammers' activeness and the challenges associated with their detection. By examining a real dataset from a large forum, we have provided insights into the prevalence of inactive spammers, which were largely overlooked as previous studies primarily focused on active spammers. Our findings emphasize the importance of considering spammers' activeness and highlight the need for caution when applying existing detection models developed mainly for active spammers. The insights gained from this research may shed light on the broader landscape of spam detection and underscore the significance of adapting detection techniques to encompass both active and inactive spammers.

Although our primary focus in this study was on political spammers, it is worth noting that the methodology and approach presented can be extended to address other types of spammers, e.g., commercial spam. The underlying principles and techniques -- incorporating social information into the model -- can be readily applied to different domains, enabling the detection and mitigation of spam in various contexts. This versatility enhances the practical applicability of our research and provides a foundation for developing effective detection mechanisms in other domains related to spamming.

Future investigations could explore additional dimensions of spammer behavior, such as the temporal dynamics of their activities or the evolving strategies employed by different spammers. Such endeavors will contribute to a more comprehensive understanding of spamming phenomena and facilitate the development of robust and adaptive detection methods to counteract the ever-evolving landscape of spam.

\bibliographystyle{unsrtnat}
\bibliography{ref}  %

\end{document}

%% file: my_package.tex
\usepackage[ruled,linesnumbered]{algorithm2e}
\usepackage{amsmath}
\usepackage{bm}
\usepackage{booktabs} %
\usepackage{color, soul}
\usepackage{graphicx}
\usepackage{makecell}
\usepackage{multirow}
\usepackage{url}
\usepackage{tabularx}
\usepackage{amsfonts}
\usepackage[T1]{fontenc}
\usepackage{algorithmic}
\usepackage[inkscapelatex=false]{svg}
\usepackage{mathtools}
\usepackage{nth}

\SetCommentSty{mycommfont}

\newcommand{\figwidth}[4]{
    \begin{figure}[tb]\centering
    \includegraphics[width=#4]{fig/#1}
    \caption{#2}
    \label{#3}\end{figure}
    {}}

\usepackage{makecell}